# Spectrum Structure of Fermion on Bloch Branes with Two Scalar-fermion Couplings

**Qun-Ying Xie**[a,b‡], **Heng Guo**[c], **Zhen-Hua Zhao**[d], **Yun-Zhi Du**[e], **Yu-Peng Zhang**[b]

[a]School of Information Science and Engineering, Lanzhou University, Lanzhou 730000, People's Republic of China
[b]Institute of Theoretical Physics, Lanzhou University, Lanzhou 730000, People's Republic of China
[c]School of Physics and Optoelectronic Engineering, Xidian University, Xi'an 710071, People's Republic of China
[d]Department of Applied Physics, Shandong University of Science and Technology, Qingdao 266590, People's Republic of China
[e]Department of Physics, Shanxi Datong University, Datong 037009, People's Republic of China

**Abstract.** It is known that the Bloch brane is generated by an odd scalar field $\phi$ and an even one $\chi$. In order to localize a bulk fermion on the Bloch brane, the coupling between the fermion and scalars should be introduced. There are two localization mechanisms in the literature, the Yukawa coupling $-\eta\bar{\Psi}F_1(\phi,\chi)\Psi$ and non-Yukawa coupling $\lambda\bar{\Psi}\Gamma^M\partial_M F_2(\phi,\chi)\gamma^5\Psi$. The Yukawa coupling has been considered. In this paper, we consider both couplings between the fermion and the scalars with $F_1 = \chi^m\phi^{2p+1}$ and $F_2 = \chi^n\phi^{2q}$, and investigate the localization and spectrum structure of the fermion on the Bloch brane. It is found that the left-handed fermion zero mode can be localized on the Bloch brane under some conditions, and the effective potentials have rich structure and may be volcano-like, finite square well-like, and infinite potentials. As a result, the spectrum consists of a series of resonant Kaluza-Klein fermions, finite or infinite numbers of bound Kaluza-Klein fermions. Especially, we find a new feature of the introduction of both couplings: the spectrum for the case of finite square well-like potentials contains discrete quasi-localized and localized massive KK modes simultaneously.

‡ E-mail: xieqy@lzu.edu.cn(corresponding author)

## 1. Introduction

The idea of extra dimensions has received increasing interest since the Arkani-Hamed-Dimopoulos-Dvali (ADD)[1] brane model and the Randall-Sundrum (RS) [2, 3] brane models (include the RS1 [2] and RS2 [3] models) were presented. Especially, the effective gravity can be still four-dimensional even for the case of the infinite extra dimensions, namely, the extra dimensions are also invisible. This surprising idea was first proposed in the RS2 model [3]. The key point is that extra dimensions are warped and hence the four-dimensional massless gravitons are trapped on the brane by an effective potential caused by the warped geometry [3]. In the RS models, our four-dimensional Universe is a thin brane, and it is assumed that all matter fields are localized on the brane [2, 3]. The generalization of the RS brane models including thick brane models can be found for examples in Refs. [4, 5, 6, 7, 8, 9]. For the review of thick branes see Ref. [10].

In this paper, our interest is focused on the five-dimensional thick brane models. In these models, one usually introduces one or more background scalar fields that play the role of generating the thick brane [11, 12, 13, 14, 15, 16]. A feature of these models is that the brane may have inner structure. In thick brane scenario, all fields including matter fields are in the bulk in principle, and our four-dimensional matter fields in the standard model are in fact the zero modes of these fundamental fields. Therefore, localization and mass spectra of various bulk matter fields especially fermions on a brane are important and interesting problems [14, 17, 18, 19, 20, 21, 22, 23, 24, 25, 26, 27, 28, 29, 30, 31, 32, 33, 34, 35, 36]. In order to localize fermion zero mode on a brane, one needs to introduce Yukawa coupling $\eta \bar{\Psi} F(\phi, \chi, \cdots) \Psi$ [17, 18, 23, 31, 32] between the bulk fermion field $\Psi$ and the background scalar fields $(\phi, \chi, \cdots)$, where $F(\phi, \chi, \cdots)$ is an odd function of the extra dimension coordinate because of the $Z_2$ symmetry of the geometry. Under this mechanism, there may exist only one bound massless fermion Kaluza-Klein (KK) mode (the fermion zero mode) [19, 21, 24], or a bound massless KK mode and finite discrete bound massive KK modes (mass gaps) [14, 20]. Furthermore, there also may exist quasi-localized fermion KK modes (also called fermion resonances) on a brane [22, 37, 38].

In Refs. [22, 39, 40], the authors investigated the localization of fermions on the Bloch brane constructed in Ref. [41]. In the Bloch brane model, there are two real background scalar fields $\phi$ and $\chi$, which are odd and even functions of the extra dimension, respectively. In order to localize the fermion zero mode on the Bloch brane, the Yukawa coupling $\eta \bar{\Psi} F(\phi, \chi) \Psi$ was introduced, where $F(\phi, \chi)$ has been taken as various forms in Refs. [22, 39, 40]. Interestingly, it was first found in Ref. [22] that there are some quasi-localized massive fermions with finite lifetime (fermion resonances) on the Bloch brane with $F(\phi, \chi) = \phi \chi$, which is related to the inner structure of the brane. Such fermion resonances were also found later in Ref. [40] with $F(\phi, \chi) = \phi$. Localization condition of the fermion zero mode for a general coupling function $F(\phi, \chi) = w_1 \phi + w_2 \chi + \phi \chi$ was analyzed in Refs. [39], where $w_1$ and $w_2$ are coupling parameters. In all these cases, the effective potentials of the fermion KK modes are volcano-like ones and the left-handed fermion zero mode can be localized on the Bloch brane under some condition. The mass spectrum is gapless and continuous because of the volcano-like potentials. Since the resonant KK modes are very different from the massive bound KK modes localized on the brane, there are many work about resonances of various fields [38, 40, 42, 43, 44, 45, 46]) after the fermion resonances were found in Ref. [22].

However, the Yukawa coupling will not work if the background scalar fields are even functions of the extra dimension, since the $Z_2$ reflection symmetry of the effective potentials for the fermion KK modes can not be ensured. Recently, in order to solve this problem, Liu *et al* presented a new localization mechanism with the coupling describing the interaction between $\pi$-meson and nucleons in quantum field theory, i.e., $\lambda \bar{\Psi} \Gamma^M \partial_M F(\phi, \chi, \cdots) \gamma^5 \Psi$ [47], which will be called as derivative coupling. This new mechanism has revealed some new interesting results [47, 46]. In the model of a two-scalar-generated brane, the fermion zero mode is localized on the brane and a mass gap between the fermion zero mode and excited KK modes was found [47], while fermion resonances were not discovered. In another dilaton-scalar-generated brane model, the fermion zero mode can also be localized on the brane but there is no mass gap [47], and fermion resonances may appear [46].

Since there are two background scalar fields, the odd scalar $\phi$ and the even one $\chi$, in the Bloch brane model, it is interesting to ask what will happen to localization and mass spectrum for fermions if we introduce both the Yukawa coupling and derivative coupling? Is the mass spectrum of fermions produced by both couplings similar to the Yukawa-coupling spectrum or the derivative coupling spectrum? Can we obtain both discrete quasi-localized and localized massive KK modes for fermions at the same time? In this paper, we would like to investigate the localization and spectrum structure of a bulk fermion on the Bloch brane by introducing both the above mentioned couplings between the fermion and the two scalars. It will be found that the effective potentials for the left- and right-handed KK fermions have three typical shapes, i.e., infinite potentials, volcano-like, and finite square well-like. Correspondingly, the spectrum looks more like the derivative coupling spectrum for the case of infinite potentials, and is similar to both the Yukawa coupling and derivative coupling spectra for the case of volcano-like potentials. Especially, for the case of finite square well-like potentials, the spectrum produced by both couplings contains discrete quasi-localized and localized massive KK modes simultaneously and so look like the sum of the Yukawa coupling spectrum (with quasi-localized modes) and derivative coupling spectrum (with localized modes), which is a new feature of the introduction of both couplings.

The paper is organized as follows. In section 2, we first give a brief review of the Bloch brane model. Then, in sections 3, 4, and 5, we study the localization and mass spectrum of fermions on the Bloch brane with only the derivative scalar-fermion coupling and both couplings. Lastly, a brief discussion and conclusion are presented in Sec. 6.

## 2. Review of the brane model

Let us review the thick branes that arise from two interactional real scalar fields $\phi$ and $\chi$ with a scalar potential $V(\phi, \chi)$. The five-dimensional action for such a system is

$$S = \int d^5x \sqrt{-g} \left[ \frac{1}{2\kappa_5^2} R - \frac{1}{2}(\partial\phi)^2 - \frac{1}{2}(\partial\chi)^2 - V(\phi, \chi) \right], \qquad (1)$$

where $R$ is the scalar curvature and the five-dimensional gravitational coupling $\kappa_5$ is set to $\kappa_5^2 = 2$ in this paper. The line-element for a static flat brane system that we are interested in is

$$ds^2 = \mathrm{e}^{2A(y)} \eta_{\mu\nu} dx^\mu dx^\nu + dy^2, \qquad (2)$$

where $y$ denotes the extra dimension and $\mathrm{e}^{2A(y)}$ is the warp factor. One can get the conformal flat metric

$$ds^2 = \mathrm{e}^{2A(z)} \left( \eta_{\mu\nu} dx^\mu dx^\nu + dz^2 \right) \qquad (3)$$

by using the following coordinate transformation

$$dz = e^{-A(y)} dy. \qquad (4)$$

The scalar fields $\phi$ and $\chi$ are functions of $y$ only for the static flat brane, i.e., $\phi = \phi(y)$ and $\chi = \chi(y)$. Substituting the line-element (2) into the five-dimensional action (1), one gets the following differential field equations

$$
\begin{aligned}
\phi'' + 4A'\phi' &= \partial_\phi V(\phi, \chi), \\
\chi'' + 4A'\chi' &= \partial_\chi V(\phi, \chi), \\
12(A')^2 + 3A'' &= -4V(\phi, \chi), \\
2(\phi'^2 + \chi'^2) &= -3V(\phi, \chi).
\end{aligned}
$$

where a prime $'$ denotes the derivative with respect to the extra dimension $y$.

It is not easy to solve the above second-order field equations. However, by introducing an auxiliary superpotential $W = W(\phi, \chi)$ [11, 12, 41, 48, 49], the above field equations of the Einstein-scalar system can be reduced to the following first-order ones

$$
\begin{aligned}
\phi' &= \partial_\phi W, \\
\chi' &= \partial_\chi W, \\
3A' &= -2W.
\end{aligned}
\qquad (5)
$$

The corresponding scalar potential $V$ is

$$V = \frac{1}{2} \left[ (\partial_\phi W)^2 + (\partial_\chi W)^2 \right] - \frac{4}{3} W^2. \qquad (6)$$

Following Ref. [50], we also consider the following generalized superpotential

$$W = \phi \left[ a \left( v^2 - \frac{1}{3}\phi^2 \right) - b\chi^2 \right]. \qquad (7)$$

Then we can get the explicit expression of the scalar potential $V(\phi, \chi)$. From Ref. [40], we know that there are two minima (two vacua) located at $\chi = 0$ and $\phi = \pm v$ in the scalar potential $V(\phi, \chi)$. For $a = 3b$, there is a very simple brane solution [50]:

$$
\begin{aligned}
\phi(y) &= v \tanh(\bar{y}), \\
\chi(y) &= v \, \mathtt{sech}(\bar{y}), \\
A(y) &= \frac{2}{3} v^2 \ln \mathtt{sech}(\bar{y}),
\end{aligned}
\tag{8}
$$

where $k = 2bv$ and $\bar{y} \equiv ky$. For $a = b$, there exists a degenerate Bloch brane solution [50]:

$$
\begin{aligned}
\phi(y) &= \frac{uv \sinh(\bar{y})}{u \cosh(\bar{y}) - c_0}, \\
\chi(y) &= \frac{2v}{u \cosh(\bar{y}) - c_0}, \\
e^{2A(y)} &= \left( \frac{c_0 - u}{c_0 - u \cosh(\bar{y})} \right)^{4v^2/9} \\
&\quad \times \exp\left[ \frac{4uv^2}{9} \left( \frac{u - c_0 \cosh(\bar{y})}{(c_0 - u \cosh(\bar{y}))^2} - \frac{1}{u - c_0} \right) \right].
\end{aligned}
\tag{9}
$$

where

$$
u = \sqrt{c_0^2 - 4},
\tag{10}
$$

and the parameter $c_0 < -2$.

The thickness of the single and double branes could be estimated as

$$
\delta \simeq
\begin{cases}
\frac{1}{bv} & \texttt{for solution (8)} \\
\frac{1}{bv} \ln \frac{-2c_0}{\sqrt{c_0^2 - 4}} & \texttt{for solution (9) with } c_0 \to -2
\end{cases}
\tag{11}
$$

The single brane locates near $z = 0$, while the double branes locate near $z = \pm \delta/2$. In what follows, we mainly discuss the case of the double brane solution (9) (i.e., $c_0 \to -2$ or $u \to 0$). Because when $u \to 0$, a little change to $c_0$ will lead to a large change of the brane thickness $\delta$, this will add difficulty for us to adjust the brane thickness, so we define a new parameter $\delta_0$:

$$
\delta_0 = \ln(4/u).
\tag{12}
$$

Now $u \to 0$ corresponds to $\delta_0 \gg 1$. The thickness of the double brane for the solution (9) is

$$
\delta \approx \frac{1}{bv} \ln \frac{4}{u} = \frac{\delta_0}{bv}.
\tag{13}
$$

Thus, the thickness of the brane can be easily adjusted by $\delta_0$, which is an integral parameter independent of the scalar potential $V(\phi, \chi)$.

## 3. Localization mechanism of a bulk fermion on the Bloch brane

In this section, we would like to investigate the localization of a bulk fermion on the Bloch brane by two fermion localization mechanisms, one is Yukawa coupling $\eta \bar{\Psi} F_1(\phi, \chi) \Psi$, the other is the derivative coupling $\lambda \bar{\Psi} \Gamma^M \partial_M F_2(\phi, \chi) \gamma^5 \Psi$ [47]. The action is given by

$$S_F = \int d^5 x \sqrt{-g} \Big( \bar{\Psi} \Gamma^M (\partial_M + \omega_M) \Psi - \eta \bar{\Psi} F_1(\phi, \chi) \Psi$$
$$+ \lambda \bar{\Psi} \Gamma^M \partial_M F_2(\phi, \chi) \gamma^5 \Psi \Big), \tag{14}$$

where $\eta$ and $\lambda$ are the scalar-fermion coupling parameters. With the metric (2), the non-vanishing components of the spin connection $\omega_M$ are $\omega_\mu = \frac{1}{2}(\partial_z A)\gamma_\mu \gamma_5$, and the five-dimensional Dirac equation reads as

$$\left[ \gamma^\mu \partial_\mu + \gamma^5 \left( \partial_z + 2A'(z) \right) - \eta e^A F_1(\phi, \chi) + \lambda F_2'(\phi, \chi) \right] \Psi = 0. \tag{15}$$

From now on, a prime stands for the derivative with respect to the conformal coordinate $z$.

Next we will study the above five-dimensional Dirac equation. Because of the chiral symmetry of a bulk fermion, naturally, we choose the following chiral decomposition:

$$\Psi(x, z) = \sum_n \Big( \psi_{Ln}(x) \hat{f}_{Ln}(z) + \psi_{Rn}(x) \hat{f}_{Rn}(z) \Big)$$
$$= e^{-2A} \sum_n \Big( \psi_{Ln}(x) f_{Ln}(z) + \psi_{Rn}(x) f_{Rn}(z) \Big), \tag{16}$$

where $\hat{f}_{L,R}(z) = e^{-2A} f_{L,R}(z)$, $\psi_{Ln,Rn}(x) = \mp \gamma^5 \psi_{Ln,Rn}(x)$. It can be shown that the four-dimensional Dirac fields $\psi_{Ln}(x)$ and $\psi_{Rn}(x)$ satisfy the four-dimensional Dirac equations $\gamma^\mu \partial_\mu \psi_{Ln,Rn}(x) = m_n \psi_{Rn,Ln}(x)$, and the KK modes satisfy the following coupled equations

$$\left[ \partial_z + \eta e^A F_1(\phi, \chi) - \lambda \, \partial_z F_2(\phi, \chi) \right] f_{Ln}(z) = + m_n f_{Rn}(z), \tag{17}$$
$$\left[ \partial_z - \eta e^A F_1(\phi, \chi) + \lambda \, \partial_z F_2(\phi, \chi) \right] f_{Rn}(z) = - m_n f_{Ln}(z). \tag{18}$$

The above coupled equations can be racast into the following Schrödinger-like equations

$$\left( - \partial_z^2 + V_L(z) \right) f_{Ln} = m_n^2 f_{Ln}, \tag{19}$$
$$\left( - \partial_z^2 + V_R(z) \right) f_{Rn} = m_n^2 f_{Rn}, \tag{20}$$

where the effective potentials for the left- and right-handed KK modes are determined by [47]

$$V_L(z) = \left( \eta e^A F_1 \right)^2 - \eta e^A \left( A' F_1 + F_1' \right) - 2\eta\lambda A F_1 F_2' + \lambda^2 \left( F_2' \right)^2 + \lambda \, F_2'', \tag{21}$$
$$V_R(z) = V_L(z)|_{\lambda \to -\lambda, \eta \to -\eta}. \tag{22}$$

In order to obtain the standard four-dimensional action for a massless and a series of massive fermions:

$$S_F = \sum_n \int d^4 x \; \bar{\psi}_n (\gamma^\mu \partial_\mu - m_n) \psi_n, \tag{23}$$

the following orthonormality conditions for $f_{L_n}$ and $f_{R_n}$ are needed:

$$\int_{-\infty}^{\infty} f_{Lm} f_{Ln} dz = \delta_{mn} = \int_{-\infty}^{\infty} f_{Rm} f_{Rn} dz, \quad \int_{-\infty}^{\infty} f_{Lm} f_{Rn} dz = 0. \tag{24}$$

From Eqs. (21) and (22), it can be seen that, if there is no coupling, i.e., $\lambda = \eta = 0$, then the effective potentials $V_L(z)$ and $V_R(z)$ will vanish and both the left- and right-handed fermions cannot be localized on the Bloch brane. Since Ref. [40] has already considered the case of the Yukawa coupling ($\eta \neq 0$ and $\lambda = 0$), in this paper, we will consider first the derivative coupling, i.e., $\eta = 0$ and $\lambda \neq 0$, by taking the simple choice $F_2(\phi, \chi) = \chi^n \phi^{2q}$, then we will come to the case of both couplings with $F_1(\phi, \chi) = \chi^m \phi^{2p+1}$ and $F_2(\phi, \chi) = \chi^n \phi^{2q}$. Note that $F_1(\phi, \chi)$ and $F_2(\phi, \chi)$ are respectively odd and even functions of the extra dimension $y$ or $z$.

## 4. The derivative coupling

Firstly, we will focus on the derivative coupling by setting $\eta = 0$. Considering the conformal transformation between the physical coordinate $y$ and conformal coordinate $z$ in Eq. (4), we rewrite the effective potentials $V_{L,R}$ (21) and (22) in the physical coordinate $y$:

$$V_L(y) = \lambda e^{2A} [\lambda F_2'^2 + A' F_2' + F_2''], \tag{25}$$

$$V_R(y) = V_L|_{\lambda \to -\lambda}, \tag{26}$$

By setting

$$F_2(\phi, \chi) = \chi^n \phi^{2q}, \tag{27}$$

the explicit forms of the effective potentials (25) and (26) for the solution (8) read as

$$\begin{aligned} V_L(y) = \frac{1}{3}\lambda k^2 v^{n+2q} \mathtt{sech}^{n+4v^2/3} \tanh^{2q}(\bar{y}) \Big[ n\left(3n + 2v^2\right) \\ + 6q(2q-1)\sinh^{-2}(\bar{y}) - (n+2q)\left(3+3n+6q+2v^2\right)\mathtt{sech}^2(\bar{y}) \\ + 3\lambda v^{n+2q}(n+4q-n\cosh(2\bar{y}))^2 \sinh^{-2}(2\bar{y})\mathtt{sech}^n(\bar{y})\tanh^{2q}(\bar{y}) \Big], \end{aligned} \tag{28}$$

$$V_R(y) = V_L|_{\lambda \to -\lambda}. \tag{29}$$

The values of the effective potentials at $y = 0$ are

$$V_L(0) = -V_R(0) = \begin{cases} -n\lambda k^2 v^n, & q = 0 \\ 2\lambda k^2 v^{n+2}, & q = 1 \\ 0, & q \geq 2 \\ \infty, & q < 0 \end{cases}. \tag{30}$$

Here, we do not consider the case of $q < 0$ since the effective potentials are singular at $y = 0$. In this paper, without loss of generality, we assume $b > 0$ and $v > 0$. The asymptotic behaviors of the effective potentials at $y \to \pm\infty$ for different values of $n$ are discussed as follows:

- When $n > -2v^2/3$ and $n \neq 0$, the effective potentials tend to vanish at the boundary and one of them is the volcano-like potential. For such potential, there may exist a massless mode and a series of massive resonant fermion modes on the brane.

- When $n = -2v^2/3$, the effective potentials have potential wells and tend to a positive constant $\frac{4}{9}\lambda^2 k^2 v^{4(1+q-v^2/3)}$ when $y \to \pm\infty$, which are finite square well-like potentials. In this case, there may be a localized fermion zero mode and a series of bound massive fermion modes.

- When $n < -2v^2/3$, the asymptotic behavior of the potentials are $V_{L,R}(\pm\infty) \to +\infty$, which will result in infinite bound massive fermion modes on the brane.

For the solution (9), the effective potentials read as

$$
\begin{aligned}
V_L(y) = {}& e^{n\delta_0 - \frac{8}{9}v^2\left(\frac{1}{2+r} - \frac{2+r\cosh(\bar{y})}{(r+2\cosh(\bar{y}))^2}\right)} \times \frac{8\lambda n b^2 v^{2+n}(2+r)^{\frac{4v^2}{9}}}{9(r+2\cosh(\bar{y}))^{4+n+\frac{4v^2}{9}}} \\
& \times \Big[ \sinh^2(\bar{y}) \left(18n\lambda v^n e^{n\delta_0}(r+2\cosh(\bar{y}))^{2-n}\right. \\
& \left. + 8v^2(5 + 3r\cosh(\bar{y}) + \cosh(2\bar{y}))\right) \\
& - 9(r+2\cosh(\bar{y}))^2 \left(2 + n + r\cosh(\bar{y}) - n\cosh(2\bar{y})\right) \Big],
\end{aligned}
\tag{31}
$$

$$
V_R(y) = V_L|_{\lambda \to -\lambda},
\tag{32}
$$

where $r = \sqrt{4 + e^{2\delta_0}}$. Next, we take research on the values of the effective potentials at $y = 0$ for different parameter $q$

$$
V_L(0) = -V_R(0) = \begin{cases}
-2n\lambda k^2 v^n e^{n\delta_0} (2+r)^{-1-n}, & q = 0 \\
8\lambda k^2 v^{2+n} e^{n\delta_0} (2+r)^{-2-n}, & q = 1 \\
0, & q \geq 2 \\
\infty, & q < 0
\end{cases}.
\tag{33}
$$

Similarly, there are also three asymptotic behaviors of the effective potential (31) for different values of $n$: $n > -2v^2/9$, $n = -2v^2/9$, and $n < -2v^2/9$ relate to volcano-like, finite square well-like, and infinite potentials, respectively. The asymptotic value of the finite square well-like potentials at the boundary is

$$
V_{L,R}(\pm\infty) = n^2\lambda^2 k^2 v^{2(n+2q)}(2+r)^{-2n} e^{2n(\frac{2}{2+r} + \delta_0)} \equiv C_\infty > 0.
\tag{34}
$$

In conclusion, the effective potentials have abundant inner structure with different values of $n$ for both brane solutions (8) and (9). The shapes of the effective potentials for the left- and right-handed fermions with different values of $n$ are shown in Figs. 1 and 2 for the brane solutions (8) and (9), respectively. The effect of $q$ on the effective potential $V_L$ is shown in Fig. 3. And Figs. 4 and 5 show how the parameters $\lambda$ and $\delta_0$ affect the effective potentials: they controls the depth and width of the effective potentials, respectively. For large enough $\delta_0$ the potential $V_R$ will have two wells.

### 4.1. The zero mode

Now, let us investigate the localization of the fermion zero mode under the derivative coupling mechanism. By setting $m_0 = 0$ and $\eta = 0$ in Eq. (17), we get the solution of the fermion zero modes

$$
f_{L0,R0}(z(y)) \propto e^{\pm\lambda F_2(\phi,\chi)} = \exp\left[\pm\lambda v^{n+2q}\mathtt{sech}^n(\bar{y})\tanh^{2q}(\bar{y})\right]
\tag{35}
$$

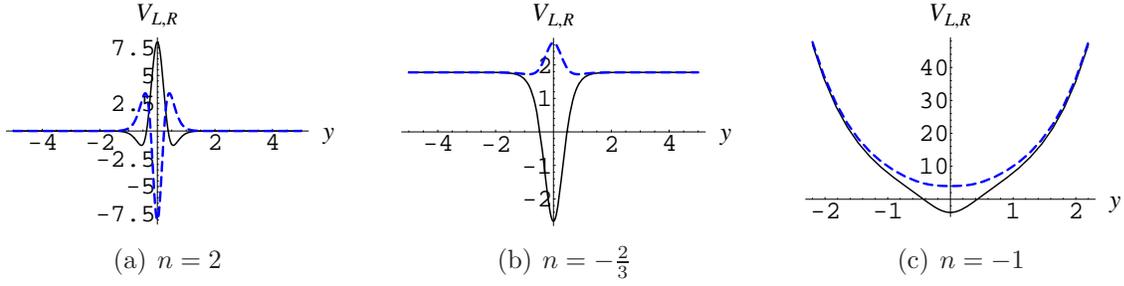

(a) $n = 2$　　(b) $n = -\frac{2}{3}$　　(c) $n = -1$

**Figure 1.** Plot of the potentials $V_L$ (black solid line) and $V_R$ (blue dashed line) for the brane solution (8) for different values of $n$ in $y$ coordinate. The parameters are set to $k = 2$, $v = 1$, $\lambda = -1$.

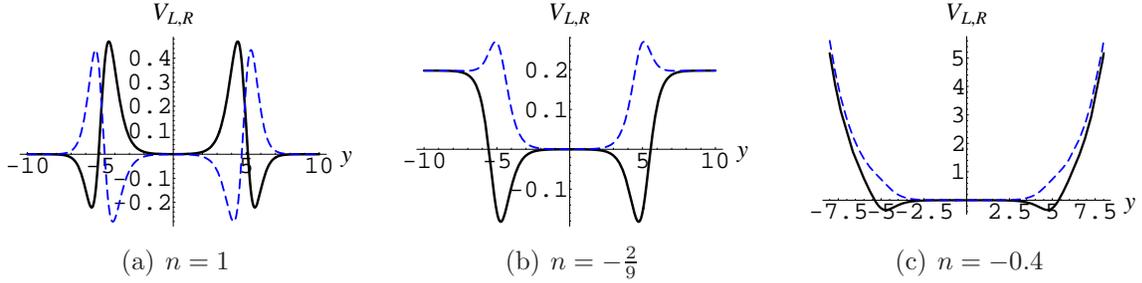

(a) $n = 1$　　(b) $n = -\frac{2}{9}$　　(c) $n = -0.4$

**Figure 2.** Plot of the potentials $V_L$ (black solid line) and $V_R$ (blue dashed line) for the brane solution (9) for different values of $n$ in $y$ coordinate. The parameters are set to $k = 2$, $v = 1$, $\lambda = -1$, $\delta_0 = 10$ ($u = 0.00018$).

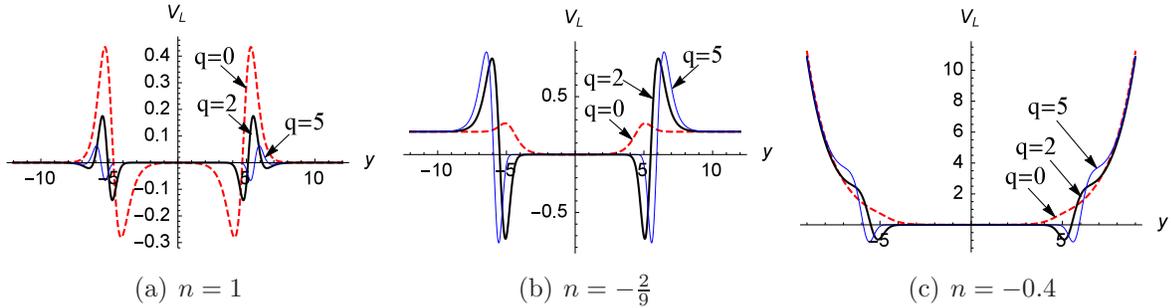

(a) $n = 1$　　(b) $n = -\frac{2}{9}$　　(c) $n = -0.4$

**Figure 3.** Plot of the potential $V_L$ for the brane solution (9) for different values of $n$ and $q$ in $y$ coordinate. The parameters are set to $k = 2$, $v = 1$, $\lambda = -1$, $\delta_0 = 10$ ($u = 0.00018$).

for solution (8) and

$$f_{L0,R0}(z(y)) \propto \exp\left[\pm 2^n \lambda u^q v^{n+2q} \sinh^{2q}(\bar{y})(u \cosh(\bar{y}) - c_0)^{-n-2q}\right] \qquad (36)$$

for solution (9).

In order to localize the zero modes on the brane, we need to consider the following

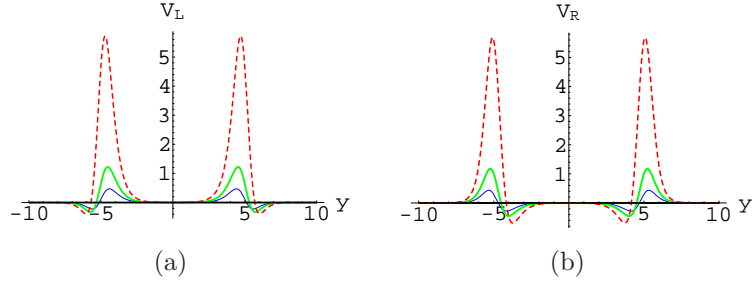

**Figure 4.** Plot of the potentials $V_{L,R}$ for the brane solution (9) for different values of $\lambda$ in $y$ coordinate. The parameters are set to $k = 2$, $v = 1$, $n = 1$, $\delta_0 = 10$, $\lambda = -1$ for blue thin lines, $\lambda = -2$ for green thick lines, and $\lambda = -5$ for the red dashed lines.

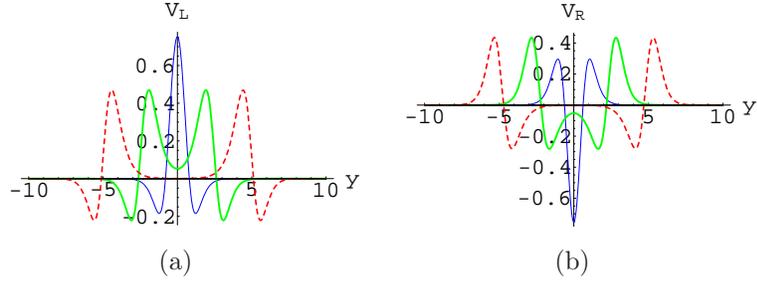

**Figure 5.** Plot of the potentials $V_{L,R}$ for the brane solution (9) for different values of $\delta_0$ in $y$ coordinate. The parameters are set to $k = 2$, $v = 1$, $n = 1$, $\lambda = -1$, $\delta_0 = 1$ for blue thin lines, $\delta_0 = 5$ for green thick lines, and $\delta_0 = 10$ for red dashed lines.

normalization condition:

$$\int dz\, f_{L0,R0}^2(z) \propto \int dz\, e^{\pm 2\lambda F_2(\phi,\chi)} < \infty, \tag{37}$$

which indicates that only one of the left- and right-handed fermion zero modes may be localized on the brane. Since we do not have the analytic expressions of the functions $A(z)$ and $\chi(z)$ in the $z$ coordinate, we have to deal with them in the $y$ coordinate. With the relation (4) between $y$ and $z$, the normalization condition of the left-handed fermion zero mode is turned out to be

$$\int_{-\infty}^{\infty} dy\, e^{2\lambda v^{n+2q}\mathtt{sech}^n(\bar{y})\tanh^{2q}(\bar{y}) - A(y)} < \infty \qquad \mathtt{for\ solution}\ (8), \tag{38}$$

$$\int_{-\infty}^{\infty} dy\, e^{2^{n+1}\lambda u^q v^{n+2q}\sinh^{2q}(\bar{y})(u\cosh(\bar{y}) - c_0)^{-n-2q} - A(y)} < \infty \qquad \mathtt{for}\ (9). \tag{39}$$

Since we have let $b > 0$ and $v > 0$, the normalization condition in (38) is

$$n < 0 \quad \mathtt{and}\ \lambda < 0, \quad \mathtt{for\ solution}\ (8). \tag{40}$$

For the solution (9), since $u > 0$, $v > 0$, $c_0 < -2$, the asymptotic behavior of the integrand in (39), named as `EI`, at $y \to \pm\infty$ is

$$
\texttt{EI} \to
\begin{cases}
+\infty & \texttt{for } n > 0, \\
+\infty & \texttt{for } n < 0 \texttt{ and } \lambda > 0, \\
0 & \texttt{for } n < 0 \texttt{ and } \lambda < 0.
\end{cases}
\tag{41}
$$

It is not difficult to know that the normalization condition for the left-handed fermion zero mode is

$$
n < 0 \texttt{ and } \lambda < 0, \texttt{ for solution } (9).
\tag{42}
$$

Under the conditions (40) and (42) for solutions (8) and (9), respectively, the left-handed fermion zero mode can be localized on the brane.

### 4.2. Volcano-like potentials and massive resonant KK modes

Next, we would like to investigate the massive fermion KK modes. We will consider three kinds of effective potentials mentioned before respectively. This subsection deals with the volcano-like potentials, for which the massive KK modes can propagate along the extra dimension and there may be some quasi-localized or resonant fermion KK modes localized on the brane within finite time [22, 38]. We will take our focus on the study of these special KK modes.

The study of resonant fermions on thick branes [12, 22, 27, 51, 52, 53, 54] is rich and interesting. In Ref. [22], the resonant fermion problem on the Bloch brane model has been studied for the first time with the Yukawa coupling $(F(\phi, \chi) = \phi\chi)$. By comparing the value of the wave-function at the origin, i.e., $|R(0)|$, to the amplitude of the plane wave oscillations at large distance along extra dimension, the authors of Ref. [22] have investigated the even massive fermion resonant modes.

In Refs. [40, 42], the relative probability [38] and transmission coefficient [43, 44, 45] methods were used to find fermion KK resonances. The conclusion is that the two methods are consistent with each other. In this paper, we would like to use the relative probability method.

One can define $|f(z)|^2$ as the probability for finding a massive fermion KK mode at the position $z$ along extra dimension. Within a narrow range $-z_b \leq z \leq z_b$ around the brane location, the relative probability named $P$ is defined in a box with borders $|z| = 10z_b$ as follows [38]:

$$
P(m) = \frac{\int_{-z_b}^{z_b} |f(z)|^2 dz}{\int_{-10z_b}^{10z_b} |f(z)|^2 dz}.
\tag{43}
$$

Note that when the mass square $m^2$ is much larger than the maxima of the effective potentials $V_{L,R}^{max}$, i.e., $m^2 \gg V_{L,R}^{max}$, the KK mode $f(z)$ can be approximated as a plane wave function $f(z) \propto \cos mz$ or $\sin mz$, and the corresponding probability would tend to 0.1.

It is known from the previous subsection that only one of the left- and right-handed fermion zero modes can be localized on the brane. In order to obtain the localized left-handed fermion zero mode, the conditions (40) and (42) should be satisfied. So we only consider the case of $\lambda < 0$ and $n < 0$. Next, we only consider the case of the solution (9).

For the set of parameters $k = 6, v = 3, \ n = -1, \ q = 0, \ \delta_0 = 25$ and $\lambda = -2$, we get four resonances with the mass spectrum $m_n^2 = \{0.5612, 1.2386, 2.1677, 3.3594\}$ for both the left- and right-handed fermions. Figure 6(a) shows the resonance spectrum of the right-handed fermions.

Next we consider the same set of parameters but with double coupling constant $\lambda$, i.e., $\lambda = -4$. It is found that there are eight fermion resonances (see Fig. 6(b)), and the mass spectrum is calculated as

$$m_n^2 = \{0.9319, 1.4071, 2.4778, 3.8263, 5.4279, 7.2663, 9.307, 11.5921\}. \quad (44)$$

For the third set of parameters $k = 6, v = 3, \ n = -1, \ q = 0 \ , \delta_0 = 50$ and $\lambda = -2$, i.e., the case with double width $\delta_0$ compared to the first set of parameters, we find ten fermion resonances (see Fig. 6(c)), and the mass spectrum is

$$m_n^2 = \{0.1414, 0.3173, 0.5617, 0.8733, 1.2509,$$
$$1.6954, 2.205, 2.7863, 3.4289, 4.1523\}. \quad (45)$$

The effects of the parameters $v$ and $q$ on the effective potential $V_R(z)$ of the right-handed fermion KK modes for the brane solution (9) are shown in Figs. 7 and 8. It is clear that the potential barrier for the case of $q = 1$ is larger that that of the case of $q = 0$. When $v < 1$, the potential barrier decreases with $q$. When $v = 1$, it increases first and then decreases with the increase of $q$. When $v > 1$, the potential barrier increases with $q$. Especially, when $v$ takes large values such as $v \geq 2$, the increase of the barrier with $q$ are huge. Therefore, the non-vanishing $q$ will greatly effect the mass spectrum of the fermion KK modes. For example, we find about 30 fermion resonances for the set of parameters given in Fig. 8(a), i.e., $k = 6, \ n = -1, \ \lambda = 2, \ \delta_0 = 25, \ v = 3, \ q = 1$ (we only show some resonances $f_R^{(n)}$ of the right-handed fermion KK modes in Fig. 9). Obviously, the number of resonances is much larger than the number 4 given in Fig. 6(a).

These results and further calculations show that the number of the resonant fermions increases with the absolute value of the coupling parameter $|\lambda|$, the brane width $\delta = \delta_0/(bv)$, and $q$ (for the case of $v > 1$).

### 4.3. Finite square well-like potentials and massive bound KK modes

When the parameter $n$ is taken as the special value $n = -\frac{2v^2}{9}$ for the brane solution (9), the effective potential (31) will change to the finite square well-like shape, for which there will appear some bound KK modes. Next, we will search for the bound states and investigate the effect of the parameters $\lambda$ and $\delta_0$ on the bound KK modes.

For $k = 2, \ v = 1, \ \beta = 0, \ n = -2/9, \ q = 0, \ \delta_0 = 1$ and $\lambda = -10$, we find five massive bound states for both the left and right-handed fermions, and the same mass spectrum of the massive bound states reads

$$m_n^2 = \{6.9527, 12.3648, 16.238, 18.7367, 20.1259\}. \quad (46)$$

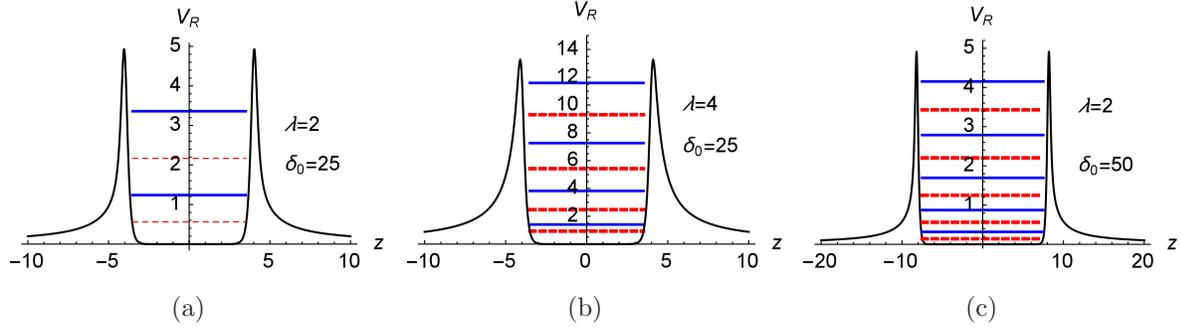

**Figure 6.** The effective potential $V_R(z)$ and the resonance spectrum $m_n^2$ of the right-handed fermions for the brane solution (9). $m_n^2$ for odd and even resonance KK modes are denoted by the red dashed and blue solid lines, respectively. The parameters are set to $k = 6$, $v = 3$, $n = -1$, $q = 0$, $\lambda$ and $\delta_0$ are shown in the figures.

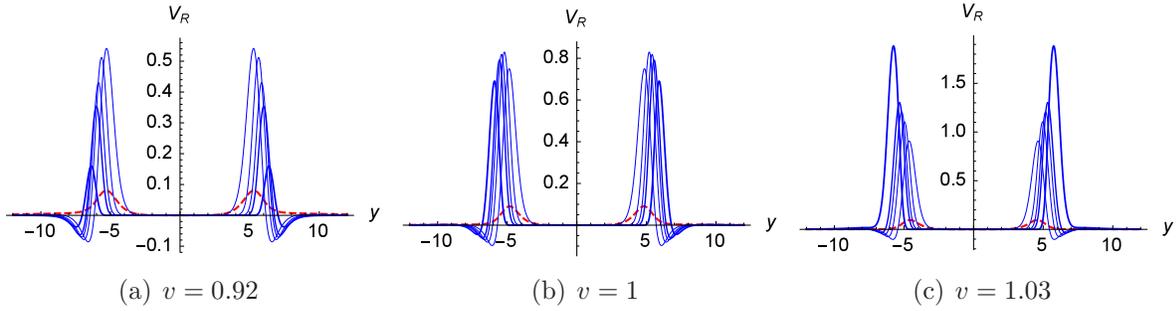

**Figure 7.** The effect of $v$ and $q$ on the effective potential $V_R(z)$ of the right-handed fermion KK modes for the brane solution (9). The parameters are set to $b = 1$, $n = -0.1$, $\lambda = 1$, $\delta_0 = 10$, $q = 0$ (red dashed line) $q = 1, 2, 3, 4, 8$ (the thickness of blue line increases with $q$), and $v = 0.92, 1, 1.03$.

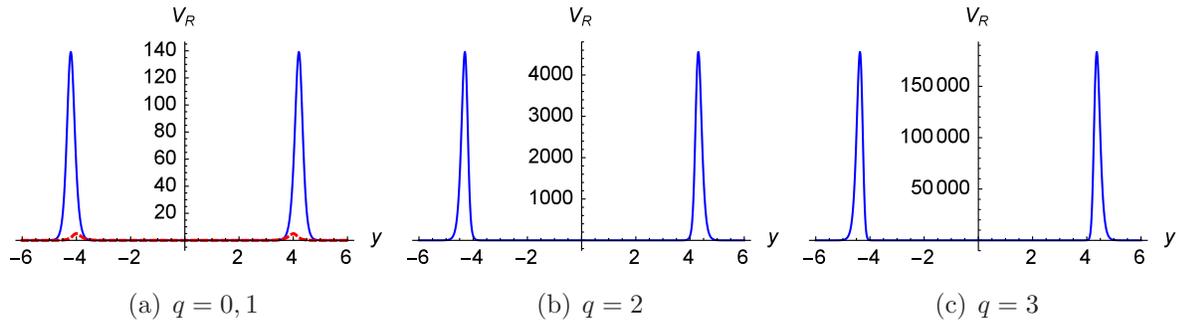

**Figure 8.** The great effect of $q$ on the effective potential $V_R(z)$ for the brane solution (9) with the large $v = 3$. The parameters are set to $k = 6$, $n = -1$, $\lambda = 2$, $\delta_0 = 25$, $q = 0$ (red dashed line), $q = 1, 2, 3$ (blue thin lines from left to right).

The finite square well-like potentials $V_{L,R}$ and the corresponding first several wave functions of the bound states are shown in Figs. 10 and 11, respectively. Comparing Fig. 10 with Fig. 11, one can get the conclusion that except for the zero mode of the left-handed fermion, both the mass spectra of the massive bound states for the left and right-handed fermions are the same.

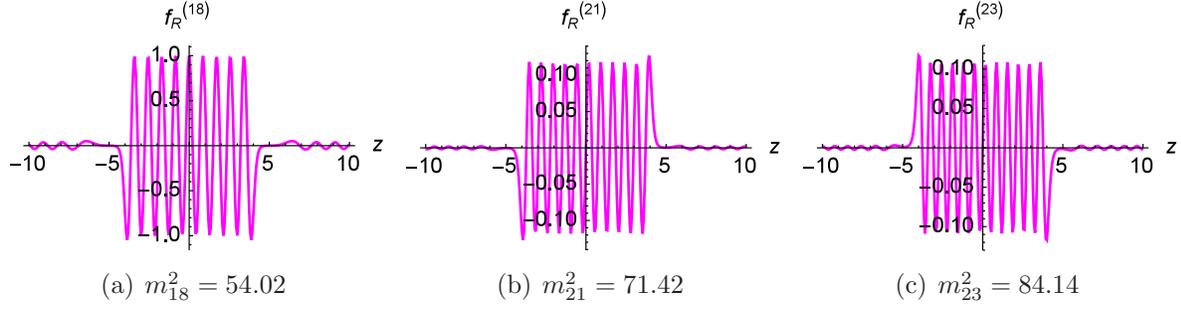

**Figure 9.** Some resonances $f_R^{(n)}$ of the right-handed fermion KK modes corresponding to the effective potential given in Fig. 8(a) with $q = 1$.

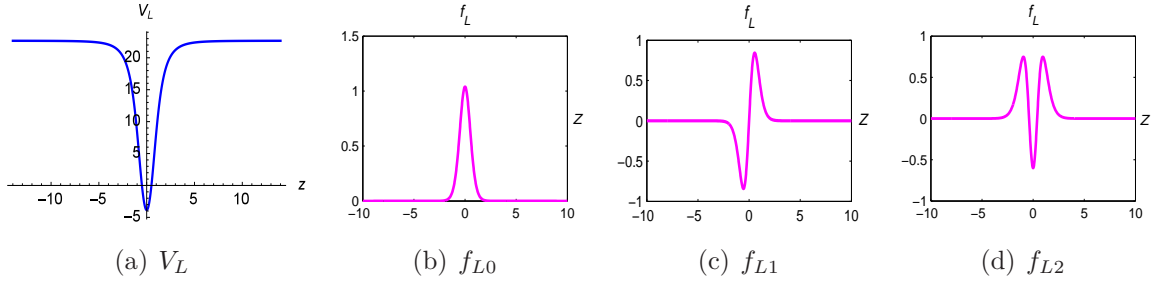

**Figure 10.** Plot of the potential $V_L$ and the wave functions of the first three bound states $f_{Ln}$ with square of mass $m_n^2$ for the brane solution (9). The parameters are set to $k = 2$, $v = 1$, $n = -2/9$, $q = 0$, $\delta_0 = 1$, $\lambda = -10$.

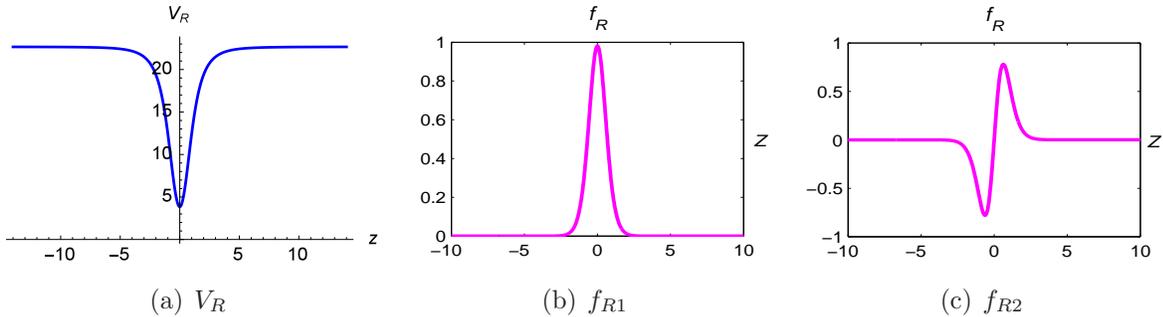

**Figure 11.** Plot of the potential $V_R$ and the wave functions of the first two bound states $f_{Rn}$ with square of mass $m_n^2$ for the brane solution (9). The parameters are set to $k = 2$, $v = 1$, $n = -2/9$, $q = 0$, $\delta_0 = 1$, $\lambda = 10$.

When double the value of the parameter $\delta_0$, i.e., $k = 2$, $v = 1$, $\beta = 0$, $n = -2/9$, $q = 0$, $\delta_0 = 2$ and $\lambda = -10$, which will cause the double width of the potential than the last set of parameters, we find five massive bound states for both the left and right-handed fermions, and the mass spectrum of the massive bound states is listed as follows:

$$m_n^2 = \{4.2592, 8.7203, 12.6784, 15.8185, 18.0067\}. \tag{47}$$

For $k = 2$, $v = 1$, $\beta = 0$, $n = -2/9$, $q = 0$, $\delta_0 = 1$ and $\lambda = -20$, double the depth of the potentials than that of the first set of parameters, we find nine bound states for the right-handed fermions, and the mass spectrum is

$$m_n^2 = \{14.6672, 27.8297, 39.4216, 49.4349,$$
$$57.9047, 64.9008, 70.5203, 74.8873, 78.1447\}. \tag{48}$$

It is interesting to note that there will appear double-well shape of the effective potentials with the increase of the width of the well, i.e, the parameter $\delta_0$. So, is there any difference between the bound states of the single- and double-well effective potentials? Next, we will find the bound states for the double-well potentials.

For $k = 6$, $v = 3$, $\beta = 0$, $n = -2$, $q = 0$, $\delta_0 = 25$, and $\lambda = -1$, we find four bound massive states for the left-handed fermions, and the mass spectrum is calculated as

$$m_n^2 = \{0.12821, 0.50687, 1.1148, 1.8859\}. \tag{49}$$

For the same set of parameters for the double-well potential, we find four bound states for the right-handed fermions:

$$m_n^2 = \{0.12817, 0.50683, 1.1148, 1.8858\}. \tag{50}$$

It can be seen that except for the left-handed zero mode, both the mass spectra for the left- and right-handed fermions are the same. This is same as the case of the single-well brane. When double the value of the parameter $\lambda$, i.e., $k = 6$, $v = 3$, $\beta = 0$, $n = -2$, $q = 0$, $\delta_0 = 25$ and $\lambda = -2$, we find seven massive bound states for both the left and right-handed fermions:

$$m_n^2 = \{0.1452, 0.5776, 1.2892, 2.2665, 3.4900, 4.9292, 6.5265\}. \tag{51}$$

We come to the conclusion that the number of the bound fermion KK modes increases with the scalar-fermion coupling parameter $\lambda$ and the brane width parameter $\delta_0$. Therefore, the double-well potentials will trap more bound KK fermions than the single-well ones.

For the infinite square well-like potentials, there exist infinite bound states. We will not discuss them here any more.

## 5. The Yukawa coupling and derivative coupling

Since the Bloch brane is generated by an odd scalar field and an even one, we should take into account both the Yukawa and derivative couplings for the localization and spectrum structure of a bulk fermion on this brane. In this subsection, we will analyze this issue.

The explicit potentials $V_{L,R}$ with the two couplings can be calculated directly according to Eqs. (21) and (22), but we do not show them here. In the following, we will just analyze the asymptotic behavior of the potentials and the localization condition for the left-handed fermion zero mode with $F_1 = \chi^m \phi^{2p+1}$ and $F_2 = \chi^n \phi^{2q}$. Since it is very complex for any $m$ and $n$, we only consider the case of $m \geq 0$.

The values of the potentials for the left and right-handed fermions at $y = 0$ are not important here, therefore, we mainly analyze the asymptotic behavior of the effective potentials at $y \to \infty$, which is given by

$$
\begin{aligned}
V_{L,R}(y \to +\infty) \to \frac{1}{9}(r+2)^{\frac{2v^2}{9}} e^{-\frac{4v^2}{9(r+2)}} \bigg\{ & (r+2)^{\frac{2v^2}{9}} e^{-\frac{4v^2}{9(r+2)}} e^{-\frac{2v^2}{9}\bar{y}} \\
& \times \Big[ k^2 n \lambda e^{n\delta} v^{n+2q} \Big( 9n\lambda e^{n\delta} v^{n+2q} e^{-2n\bar{y}} + \big( 9n + 2v^2 \big) e^{-n\bar{y}} \Big) \\
& + \eta k e^{m\delta} \big( 9m + 2v^2 \big) v^m v^{2p+1} e^{-m\bar{y}} + 9\eta^2 e^{2m\delta} v^{2m+4p+2} e^{-2m\bar{y}} \Big] \\
& - 4\eta k n \lambda e^{(m+n)\delta} v^{m+n+2p+2q+3} e^{-(m+n)\bar{y}} \bar{y} \bigg\} e^{-\frac{2v^2}{9}\bar{y}}.
\end{aligned}
\tag{52}
$$

When $n = -2v^2/9$ and $m > 0$ the potentials are constant at $y \to \pm\infty$:

$$
V_{L,R}(y \to \pm\infty) = k^2 \lambda^2 n^2 (r+2)^{\frac{4v^2}{9}} v^{2n+4q} e^{2n\delta - \frac{8v^2}{9(r+2)}},
\tag{53}
$$

which is independent of the coupling $\eta$. We dot not consider the case of $m < 0$, but we note here that the potentials are also constant at $y \to \pm\infty$ when $m = -2v^2/9$ and $n > 0$:

$$
V_{L,R}(y \to \pm\infty) = \eta^2 (r+2)^{\frac{4v^2}{9}} v^{2m+4p+2} e^{2m\delta - \frac{8v^2}{9(r+2)}}.
\tag{54}
$$

The potentials diverge at $y \to \pm\infty$ for the case of $m > 0$ and $n < -2v^2/9$. In other cases with $m \geq 0$, The potentials vanish at the boundary. So, there are three typical shapes for both couplings with nonvanishing $\eta$ and $\lambda$. The corresponding asymptotic behaviors can be summarized as follows:

$$
V_{L,R}(y \to \pm\infty) \to
\begin{cases}
+\infty & \text{for } n < -\frac{2v^2}{9}, \ m > 0 \\
C_\infty & \text{for } n = -\frac{2v^2}{9}, \ m > 0 \\
0 & \text{for other cases}
\end{cases} .
\tag{55}
$$

where $C_\infty$ is a positive constant given in Eq. (53).

In the following, we will focus on the case of $F_1 = \phi$ and $F_2 = \chi^n$. The typical shapes of the effective potential $V_L$ for only the Yukawa coupling, only the derivative coupling, and both couplings are shown in Fig. 12.

### 5.1. The zero mode

By setting $m_0 = 0$ in Eq. (17), we get the zero mode $f_{L0}(y)$ for the full coupling:

$$
\begin{aligned}
f_{L0}(y) &\propto \exp\left[ \lambda \, F_2(\phi, \chi) - \eta \int_0^z e^{A(z')} F_1(\phi, \chi) dz' \right] \\
&\propto \exp\left[ \lambda \, F_2(\phi, \chi) - \eta \int_0^y F_1(\phi, \chi) \, dy' \right].
\end{aligned}
\tag{56}
$$

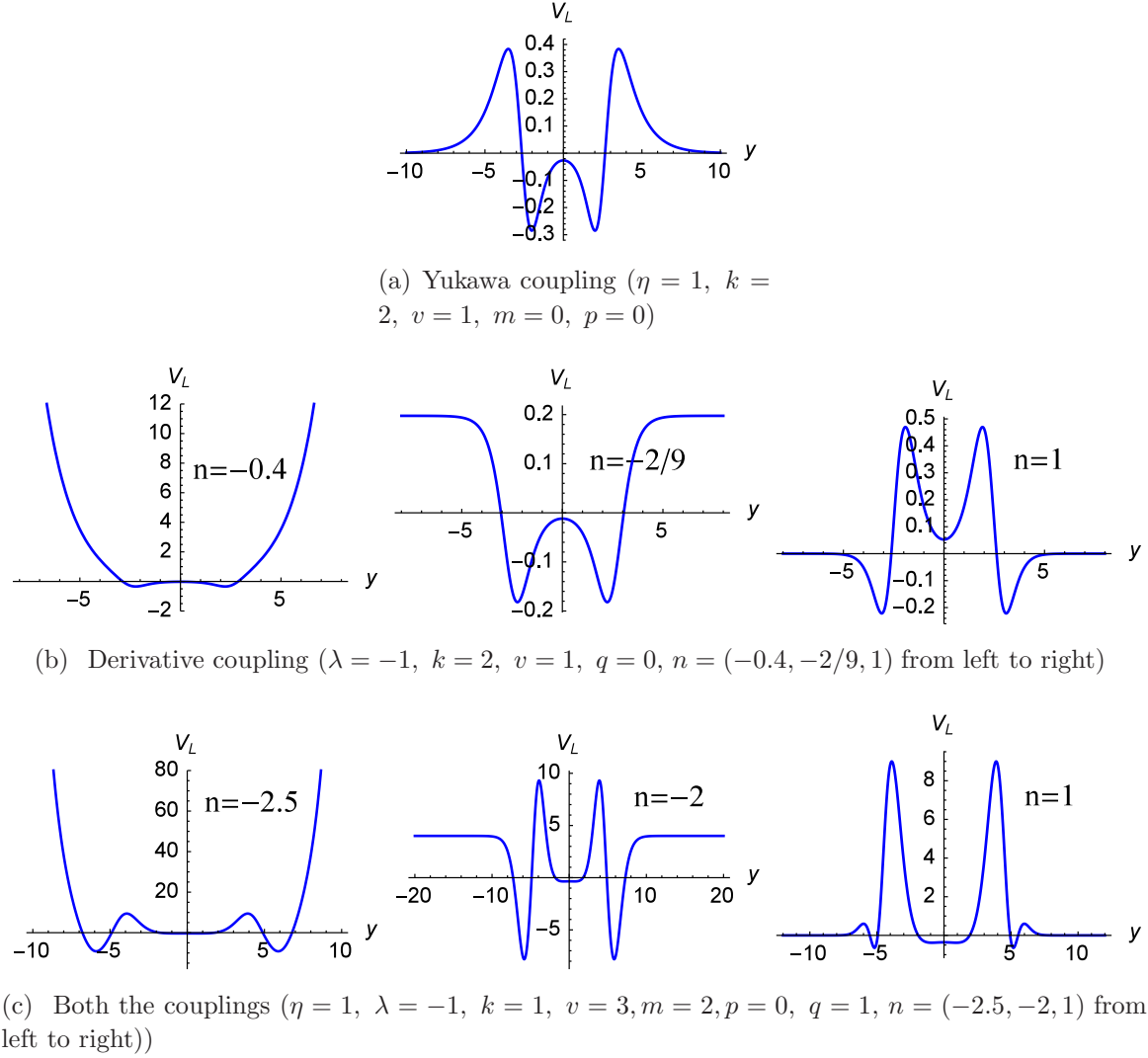

(a) Yukawa coupling ($\eta = 1,\ k = 2,\ v = 1,\ m = 0,\ p = 0$)

(b) Derivative coupling ($\lambda = -1,\ k = 2,\ v = 1,\ q = 0,\ n = (-0.4, -2/9, 1)$ from left to right)

(c) Both the couplings ($\eta = 1,\ \lambda = -1,\ k = 1,\ v = 3, m = 2, p = 0,\ q = 1,\ n = (-2.5, -2, 1)$ from left to right))

**Figure 12.** The effective potential $V_L$ for only the Yukawa coupling, only the derivative coupling, and both couplings. The parameter $\delta$ is set as $\delta = 5$.

For simplify, we will focus on the case of $F_1 = \phi$ and $F_2 = \chi^n$. The normalization condition for this zero mode is

$$\int_{-\infty}^{+\infty} dy \exp\left(2\lambda\chi^n - A(y) - 2\eta\int_0^y dy'\phi(y')\right) < \infty. \tag{57}$$

The integrand in Eq. (57) can be expressed as

$$\texttt{EI}_\texttt{F} \propto \exp\Bigg[2\lambda\Big(\frac{2v}{u\cosh(\bar{y}) - c_0}\Big)^n - \frac{2uv^2(u - c_0\cosh(\bar{y}))}{9(u\cosh(\bar{y}) - c_0)^2}$$
$$+ \Big(\frac{2v^2}{9} - \frac{\eta}{b}\Big)\ln\left(u\cosh(\bar{y}) - c_0\right)\Bigg]. \tag{58}$$

The asymptotic behavior of $\texttt{EI}_\texttt{F}$ at $y \to \pm\infty$ is

$$\texttt{EI}_\texttt{F}(y \to \pm\infty) \to \begin{cases} e^{(\frac{2v^2}{9} - \frac{\eta}{b})k|y|} & \text{for } n > 0 \\ e^{2n+1}\left(\frac{v}{u}\right)^n \lambda e^{-nk|y|} & \text{for } n < 0 \end{cases}. \tag{59}$$

So, the normalization condition is $\eta > \frac{2bv^2}{9}$ and $\lambda < 0$ for $n > 0$ and $n < 0$, respectively. It is interesting to note that the normalization condition is related to not only the coupling constant $\eta$ or $\lambda$ but also the parameter $n$.

The zero mode (56) in the $y$ coordinate can be explicitly written as

$$f_L(y) \propto [u \cosh(\bar{y}) - c_0]^{-\frac{\eta}{2b}} e^{\lambda \left( \frac{2v}{u \cosh(\bar{y}) - c_0} \right)^n}, \tag{60}$$

from which one can see that it is localized near the Bloch brane and is the ground state. Hence there is no tachyon spectrum for the KK fermions.

### 5.2. The massive resonant and bound KK modes

At last, we discuss the massive resonant and bound KK modes.

For the case of infinite potential (see the first figure in Fig. 12(c)), there exists infinite bound KK modes. For the case of volcano-like potential (see the third figure in Fig. 12(c)), one can obtain some resonant KK modes. We do not go further for these two situations. We will analyze the particular case shown in the second figure in Fig. 12(c).

For the second potential in Fig. 12(c), we find four bound KK modes and two resonant ones. In fact, the shape of the potential depends on the parameters. Here, we consider another set of parameters, $\eta = 1, \lambda = -2, k = 1, v = 1, \delta_0 = 20, m = 2, n = -2/9, p = 0, q = 1$, for which we find six bound KK modes with $m^2 = (0, 0.01, 0.02, 0.05, 0.10, 0.16)$ and four resonant ones with $m_n^2 = (0.30, 0.39, 0.49, 0.60)$ and the results are shown in Fig. 13. Compared with the case of only derivative coupling (finite square well-like potential, see Fig. 10), there are some resonant KK modes beside the bound ones. This is due to the appearance of the potential barrier in the potential well. The number of resonant KK modes will increase with the parameters $\eta$ and $m$ when $v > 1$ and decrease with $k$ and $q$. The number of bound KK modes will increase with $k$, $v$ and $\lambda$.

## 6. Discussions and conclusions

In this paper we have investigated the localization and spectrum structure of a bulk fermion on the Bloch brane. There are two background scalar fields in this brane model, an odd $\phi$ and an even $\chi$. Therefore, we considered the full coupling between the fermion and the two scalar fields, including the Yukawa coupling $-\eta \bar{\Psi} F_1(\phi, \chi) \Psi$ and the derivative coupling $\lambda \bar{\Psi} \Gamma^M \partial_M F_2(\phi, \chi) \gamma^5 \Psi$. The left-handed fermion zero mode and its localization condition are obtained. If we only consider the Yukawa coupling with $F_1 = \phi$, the effective potential $V_L$ for the left-handed KK modes is volcano-like and there are some resonant KK modes [40]. For the case of the derivative coupling with $F_2 = \chi^n \phi^{2q}$, the effective potential $V_L$ has three types of shapes: volcano-like, finite square well-like, and infinite potentials. Correspondingly, there may appear finite numbers of resonant KK fermions, finite and infinite bound KK fermions. The number of the resonant or bound KK fermions increases with the scalar-fermion coupling

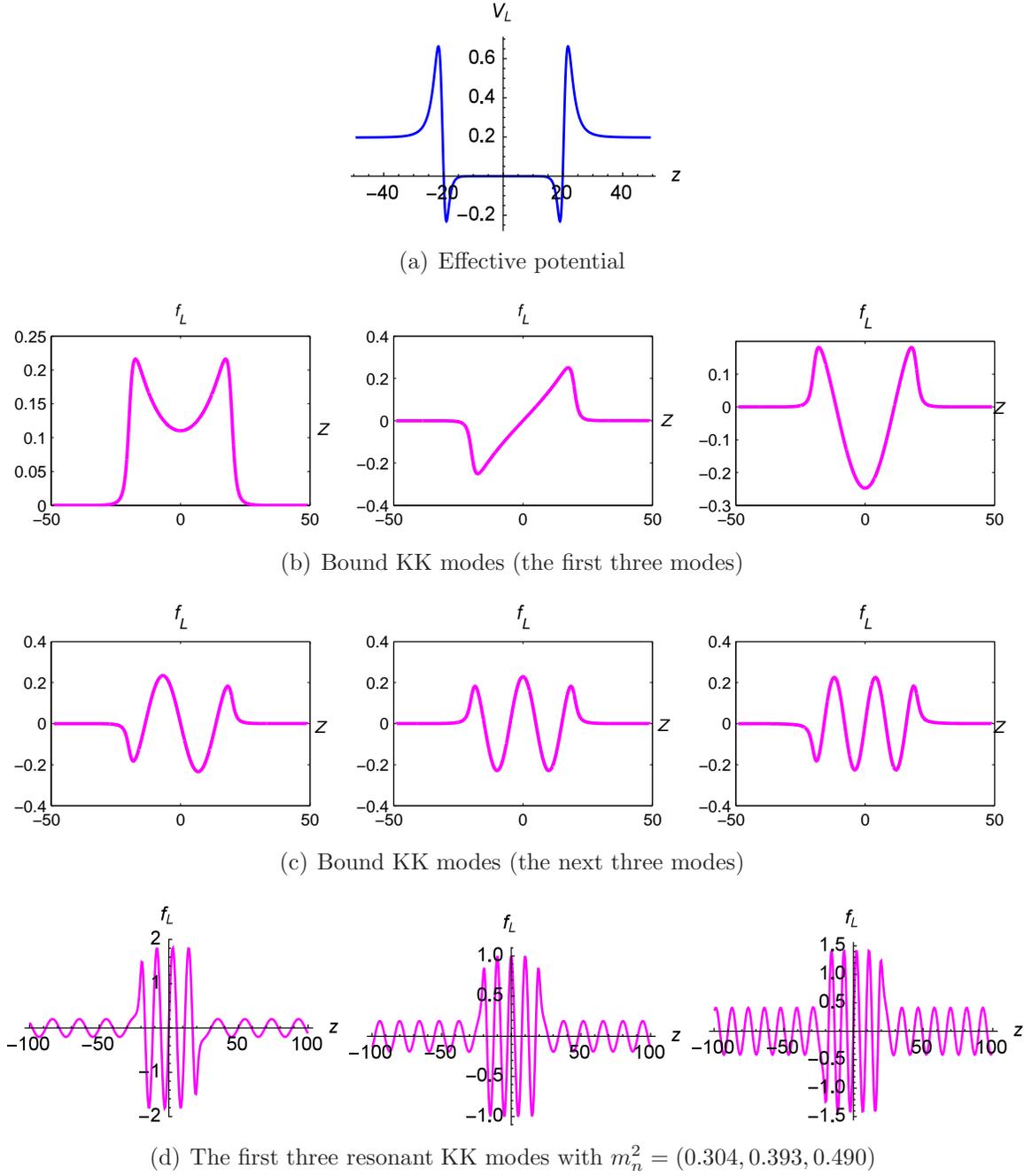

(a) Effective potential

(b) Bound KK modes (the first three modes)

(c) Bound KK modes (the next three modes)

(d) The first three resonant KK modes with $m_n^2 = (0.304, 0.393, 0.490)$

**Figure 13.** The effective potential $V_L$ and corresponding bound and resonant KK modes for the full coupling. The parameters are set to $\eta = 1, \lambda = -2, k = 1, v = 1, \delta_0 = 20, m = 2, n = -2/9, p = 0, q = 1$.

parameter $\lambda$ and the brane width parameter $\delta_0$. The similar results were obtained for the full coupling except for the case of $n = -2v^2/9$, for which the effective potential $V_L$ has not only well but also barrier, and subsequently both massive bound and resonant modes appear in the spectrum. This is a new result in the case of both couplings. The spectrum produced by the action of both couplings looks more like the derivative coupling spectrum when $n < -2v^2/9$ and is similar to both the Yukawa coupling and new coupling spectra when $n > -2v^2/9$. When $n = -2v^2/9$, it contains discrete quasi-localized and localized massive KK modes and so look like the sum of the Yukawa coupling spectrum (with quasi-localized modes) and new coupling spectrum (with localized modes).

## 7. Acknowledgement

We are graceful to the referees for their helpful comments and suggestions, which are important for the improvement of this paper. This work was supported by the National Natural Science Foundation of China (Grants No. 11522541 and No. 11375075), and the Fundamental Research Funds for the Central Universities (Grant No. lzujbky-2015-107 and No. lzujbky-2016-k04). H. Guo. was supported by the National Natural Science Foundation of China (Grant No. 11305119) and Natural Science Basic Research Plan in Shanxi Province of China (Program No. 2015JQ1015). Z.-H. Zhao was supported by the National Natural Science Foundation of China (Grant No. 11305095).